\documentclass[reprint,aps,prl,superscriptaddress,longbibliography]{revtex4-1}
\usepackage[latin9]{inputenc}
\usepackage{color}
\usepackage{times}
\usepackage{amsmath}
\usepackage{amssymb}
\usepackage{stmaryrd}
\usepackage{graphicx}
\usepackage[unicode=true,
 bookmarks=true,bookmarksnumbered=false,bookmarksopen=false,
 breaklinks=false,pdfborder={0 0 1},backref=false,colorlinks=true]
 {hyperref}
\hypersetup{
 linkcolor=magenta, urlcolor=blue, citecolor=blue, pdfstartview={FitH}, hyperfootnotes=true, unicode=true}

\makeatletter

\@ifundefined{textcolor}{}
{
 \definecolor{BLACK}{gray}{0}
 \definecolor{WHITE}{gray}{1}
 \definecolor{RED}{rgb}{1,0,0}
 \definecolor{GREEN}{rgb}{0,1,0}
 \definecolor{BLUE}{rgb}{0,0,1}
 \definecolor{CYAN}{cmyk}{1,0,0,0}
 \definecolor{MAGENTA}{cmyk}{0,1,0,0}
 \definecolor{YELLOW}{cmyk}{0,0,1,0}
}

\usepackage{amsfonts}\usepackage{tabularx}\usepackage{dcolumn}\usepackage{bm}\usepackage{graphicx}\usepackage{epstopdf}

\hypersetup{urlcolor=blue}

\usepackage{tensor}
\usepackage{braket}

\DeclareMathOperator{\sgn}{sgn}
\newcommand{\Hc}{\mathrm{H.c.}}
\renewcommand{\(}{\left(}
\renewcommand{\)}{\right)}

\usepackage[capitalise,compress]{cleveref}
\crefname{section}{Sec.}{Secs.}
\Crefname{section}{Section}{Sections}
\crefrangelabelformat{equation}{\textup{(#3#1#4)}--\textup{(#5#2#6)}}

\makeatother
\begin{document}

\title{Anomalous Quantum Information Scrambling for $\mathbb{Z}_3$ Parafermion  Chains}

\author{Shun-Yao Zhang}
\affiliation{Center for Quantum Information, IIIS, Tsinghua University, Beijing 100084, PR China}

\author{Dong-Ling Deng}
\email{dldeng@tsinghua.edu.cn}
\affiliation{Center for Quantum Information, IIIS, Tsinghua University, Beijing 100084, PR China}
\affiliation{Shanghai Qi Zhi Institute, 41th Floor, AI Tower, No. 701 Yunjin Road, Xuhui District, Shanghai 200232, China}

\begin{abstract}
Parafermions are exotic quasiparticles with non-Abelian fractional statistics that  could be exploited to realize universal topological quantum computing. Here, we study the scrambling of quantum information in one-dimensional parafermionic chains, with a focus on $\mathbb{Z}_3$ parafermions in particular. We use the generalized out-of-time-ordered correlators (OTOCs) as a measure of the information scrambling and introduce an efficient method based on matrix product operators  to compute them. With this method, we compute the OTOCs for $\mathbb{Z}_3$ parafermions chains up to $200$ sites for the entire early growth region. We find that, in stark contrast to the dynamics of conventional fermions or bosons, the  information scrambling light cones for parafermions can be both symmetric and asymmetric, even for inversion-invariant Hamiltonians involving only hopping terms. In addition, we find a deformed light cone structure with a sharp peak at the boundary of the parafermion chains in the topological regime, which gives a unambiguous evidence of the strong zero modes at infinite temperature. 
\end{abstract}

\pacs{}

\maketitle

\begin{figure}
  \centering\includegraphics[width=0.49\textwidth]{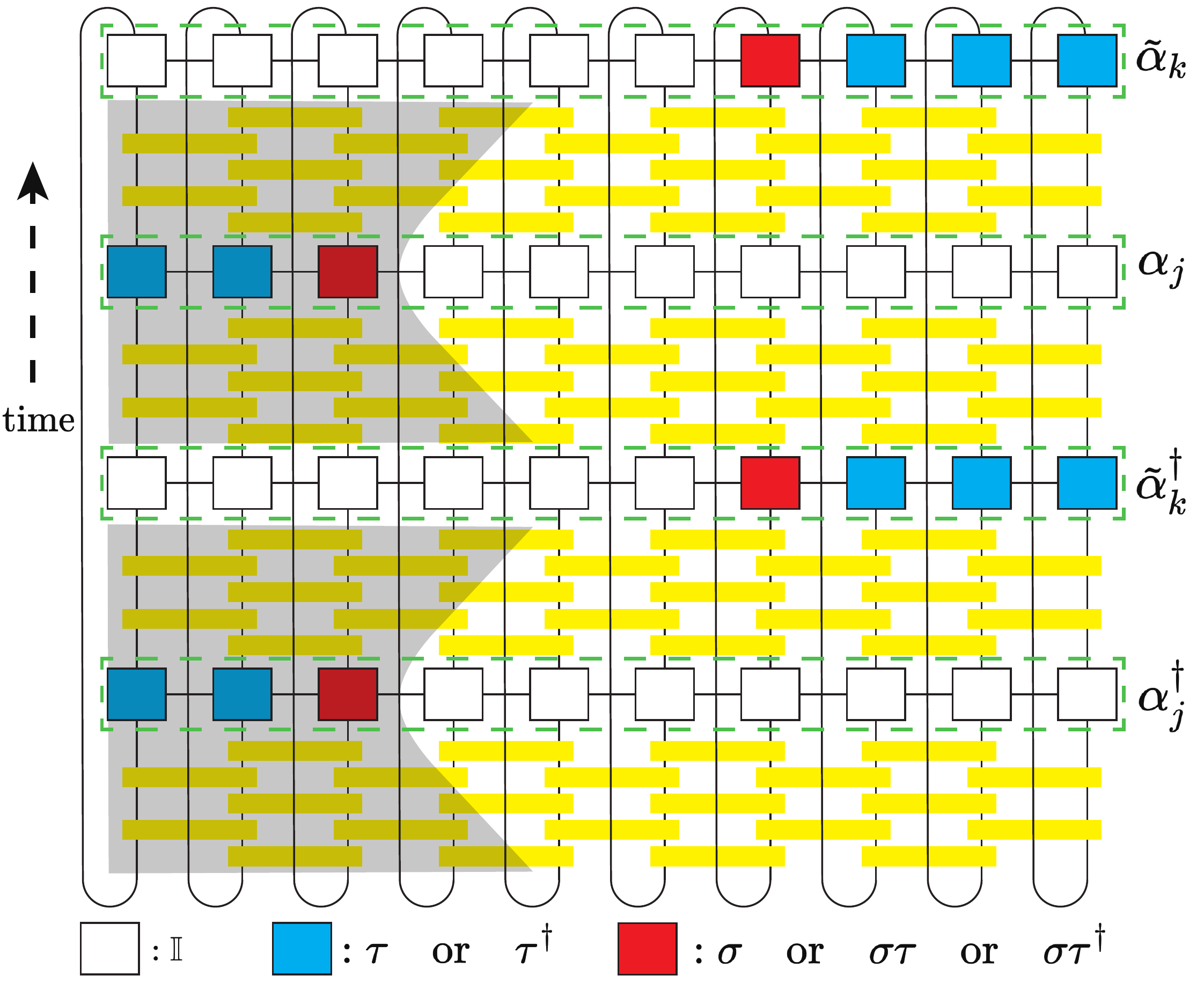}
  \caption{A schematic illustration of the matrix product operator algorithm for computing the out-of-time-ordered correlators for parafermion chains. Here, the computation of $\tilde{F}_{j,k}=\left\langle\alpha_{j}^{\dagger}(t)\tilde{\alpha}_{k}^{\dagger}(0)  \alpha_{j}(t) \tilde {\alpha}_{k}(0)  \right\rangle$  is shown.  The white (blue, red) blocks represent the local identity ($\tau$ or $\tau^\dagger$, $\sigma$ or $\sigma\tau$ or $\sigma\tau^\dagger$) operators. The yellow blocks represent the Heisenberg time-evolved gates and the shaded regions indicate the  induced light cones of $\alpha_{j}(t)$ and $\alpha_{j}^{\dagger}(t)$, respectively. For more details, see the Supplementary Materials \cite{SuppAQSPA}. 
  }
  \label{Algorithm Illustration}
\end{figure}

Non-Abelian anyons are elusive quasiparticle excitations emerged from certain topological phases of matter \cite{nayak2008non}. They obey non-Abelian braiding statistics and are the building blocks for realizing topological quantum computing \cite{nayak2008non,kitaev2003fault}.   A prominent example of non-Ableian anyons involves  parafermions \cite{clarke2013exotic,Fendley_2012,alicea2016topological, lindner2012fractionalizing,vaezi2013fractional,vaezi2014superconducting,cheng2012superconducting,barkeshli2014synthetic,
stoudenmire2015assembling,tsvelik2014integrable,klinovaja2014parafermions,zhang2014time,orth2015non,
alexandradinata2016parafermionic,alavirad2017z,calzona2018z,mazza2018nontopological,hutter2016quantum,alicea2015designer,
chew2018fermionized,zhuang2015phase,li2015criticality}, which generalize the extensively studied Majornana fermions \cite{kitaev2001unpaired,fu2008superconducting,alicea2011non} and similarly underpin a host of novel phenomena. In particular, braiding of parafermions could supply a richer set of topologically protected operations compared with the Majorana case. Although these operations are still not sufficient to enable computational universality, coupled parafermion arrays in quantum Hall architectures  can lead to Fibonacci anyons, which would then harbor universal topological quantum computation \cite{mong2014universal}. Here, we study the scrambling of quantum information in $\mathbb{Z}_3$ parafermion chains,  by introducing an efficient algorithm based on matrix product operators (MPOs) to compute the generalized out-of-time-ordered correlators (OTOCs) (see Fig. \ref{Algorithm Illustration} for an illustration).

Information scrambling in quantum many-body systems  has attracted tremendous recent attention \cite{hayden2007black,sekino2008fast,shenker2014black,hosur2016chaos,landsman2019verified}. It plays an important role in understanding a wide spectrum of elusive phenomena, ranging from the black hole information problem \cite{hayden2007black,sekino2008fast,shenker2014black,hosur2016chaos,landsman2019verified} and quantum chaos \cite{Stockmann2000Quantum} to quantum thermalization  and many-body localization \cite{Altman2018Many,Nandkishore2015Many,Abanin2019Colloquium}. Whereas black holes are conjectured to be the fastest scramblers in nature \cite{Lashkari2013Towards}, the information scrambling in a many-body localized system is much slower \cite{Swingle2017Slow,fan2017out,Huang2017Out}. For conventional bosonic or fermionic systems with translation and inversion symmetries,  information scramble in a spatially symmetric way \cite{Lauchli2008Spreading,Cheneau2012Light,Bohrdt2017Scrambling}. In sharp contrast, it has been shown that asymmetric information scrambling and particle transport could occur for Abelian anyons due to the interplay of anyonic statistics and interactions \cite{liu2018asymmetric}. In addition, asymmetric butterfly velocities in different directions have also been studied for certain spin Hamiltonians and random unitary circuits \cite{Stahl2018Asymmetric,Zhang2020Asymmetric}. Yet, despite these notable progresses, scrambling of information in systems with non-Abelian anyons still remains barely explored. A major challenge faced along this line  is that the computation of the OTOC, which is a characteristic measure of information scrambling, is notoriously difficult owing to the exponential growth of the Hilbert dimension involved. 

In this paper, we study the scrambling of information in $\mathbb{Z}_3$ parafermion chains. We mainly address two questions: (a) How to efficiently access information scrambling for parafermion chains; (b)  How information scrambles in parafermion chains? For (a), we propose an efficient algorithm based on MPOs \cite{zwolak2004mixed,verstraete2004matrix,vidal2007classical, schollwock2011density,xu2020accessing,white2018quantum,hemery2019matrix} to compute the generalized OTOCs and  demonstrate its effectiveness by computing the OTOCs for $\mathbb{Z}_3$ parafermion chains as long as $200$ sites for the entire early growth region. For (b), we find that the information scrambling light cones for parafermions can be both symmetric and asymmetric depending on the specific parameter values, even for inversion-invariant Hamiltonians involving only hopping terms. 
In addition, we find a deformed light cone with a sharp peak at the boundary of the parafermion chains in the topological region,  which provides a unambiguous evidence for the existence of strong zero modes at infinite temperature. Our results reveal some crucial aspects of information scrambling for non-Abelian anyons, which would provide a valuable guide for future studies on such exotic quasiparticles in both theory and experiment.

{\it The model Hamiltonian.---}We consider the following Hamiltonian for a  $\mathbb{Z}_3$ parafermion chain \cite{li2015criticality},  which arises naturally from coupled domain walls on the edge of two dimensional (2D) fractionalized topological insulators \cite{lindner2012fractionalizing,cheng2012superconducting,Klinovaja2014Kramers,clarke2013exotic}:
\begin{equation}
H = -t_1\sum_{j} e^{i \theta}  \alpha_{j}^\dagger\alpha_{j+1}  + t_2\sum_{j} e^{i \phi}  \alpha_{j}^\dagger\alpha_{j+2} +  \text{H.c.},
\label{para_model}
\end{equation} 
where $\alpha_j$ are parafermion operators obeying $\alpha_j^3 =1$, $ \alpha_j^\dagger = \alpha_j^2$, and commutation relations $\alpha_i\alpha_j = \alpha_j\alpha_i\omega^{\text{sgn}(j-i)} ,  \omega =e^{i{2\pi \over 3}}$,
and  $t_1$, $t_2$ control the strength of nearest-neighbor and next-nearest-neighbor hoppings, respectively. Below, we set $t_1 = 1$ as the energy unit.
By the generalized Jordan-Wigner transformation \cite{jordan1993paulische} $\alpha_{2j-1} = \left(\prod_{k=1}^{j-1} \tau_k\right) \sigma_j ,  \alpha_{2j} = \omega \left(\prod_{k=1}^{j-1}\tau_k\right)  \sigma_j\tau_j$,
the parafermion chain can be mapped to an extended $\mathbb{Z}_3$ clock model,
\begin{eqnarray}
H_c & =& -t_1 \sum_{j} e^{i\theta}\omega \sigma_j^\dagger \sigma_{j+1} - t_1 \sum_{j} e^{i\theta}\omega \tau_j \\ 
&+& t_2 \sum_{j} (e^{i\phi}\sigma_j^\dagger \tau_{j+1} \sigma_{j+1}  +  e^{i\phi}\sigma_j^\dagger \tau_{j} \sigma_{j+1}) +  \text{H.c.}, \nonumber
\label{clock_model}
\end{eqnarray}
where $\sigma_j$ and $\tau_j$ are generalized spin operators, satisfying $\sigma_j^3=\tau_j^3 =1$ , $\sigma_j \tau_j = \omega \tau_j \sigma_j$ on site and commute with each other off site.

A key quantity to measure information scrambling for parafermion chains is the generalized squared commutator of two local parafermion operators, defined as $C_{j,k}(t)=\left\langle\left[\alpha_{j}(t), \alpha_{k}\right]_\omega^{\dagger}\left[\alpha_{j}(t), \alpha_{k}\right]_\omega\right\rangle$,
which is closely related to the out-of-time-ordered correlator $F_{j,k}(t) = \left\langle\alpha_{j}^{\dagger}(t) \alpha_{k}^{\dagger}(0) \alpha_{j}(t) \alpha_{k}(0)\right\rangle\omega^{ \operatorname{sgn}(j-k)}$, 
through the relation $C_{j,k}=2[1-\operatorname{Re}(F_{j,k})]$. Here the commutator $[\alpha_{j}, \alpha_{k}]_\omega$  is defined as $[\alpha_{j}, \alpha_{k}]_\omega = \alpha_{j} \alpha_{k} -  \omega^{ \operatorname{sgn}(k-j) }\alpha_{k} \alpha_{j}$, and the average $\left \langle \cdot \right \rangle \equiv \text{Tr}(\cdot)/3^L$ is measured from the infinite-temperature ensemble. Due to the mathematically equivalence of these two models,  one can calculate the physical quantities for  the parafermion chains by using the mapped clock models. Yet, local operators in the parafermion model will become highly non-local in the mapped clock model due to the string operators in the generalized Jordan-Wigner transformation. This poses a notable challenge in computing the OTOCs for parafermions.  In the following, we introduce an efficient algorithm that could overcome this difficulty. 


{\it Algorithm.---}Our algorithm is inspired by Xu and Swingle's MPO approach to  computing OTOCs for spin systems in Ref. \cite{xu2020accessing}. Suppose we are considering the OTOC for two local Heisenberg operators $W_0$ and $V_r$ with distance $r\gg 0$, the expansion of $W_0(t)$ approximately forms a light cone,  which is confined by the Lieb-Robinson bound \cite{lieb1972finite}.  The entanglement grow massively inside the light-cone while remain vanishingly small outside. As a result, for computing OTOCs near or outside the light-cone a moderate bond dimension for MPOs suffices. In other words,  as long as the local operator $V_r$ lies outside the light-cone of $W_0(t)$,   the calculation of the OTOC using MPO is always efficient and effective.  
However,  for parafermion models,   local parafermion operators  become highly non-local string operators  under the Jordan-Wigner transformation.  For instance,  we consider the OTOC between $\alpha_{2j+1}$ and $\alpha_{2k+1}$ for parafermions,  which is equivalent to calculate the OTOC of two non-local operators $\left(\prod_{n=1}^{j-1} \tau_n\right) \sigma_{j}$ and $\left(\prod_{n=1}^{k-1} \tau_n\right) \sigma_{k}$ in the $\mathbb{Z}_3$ spin model.  These two string operators have vanishing distance between them, which renders the direct MPO approach inapplicable. 

To overcome this problem, we find that instead of $F_{j,k}$ one can calculate the equivalent quantity $\tilde{F}_{j,k}$ defined as:
\begin{equation}
\tilde{F}_{j,k}(t) = \left \{
\begin{aligned}
&\left\langle \tilde{\alpha}_{j}^{\dagger}(t) \alpha_{k}^{\dagger}(0) \tilde{\alpha}_{j}(t) \alpha_{k}(0)\right\rangle , j \geq k \\
&\left\langle\alpha_{j}^{\dagger}(t)\tilde{\alpha}_{k}^{\dagger}(0)  \alpha_{j}(t) \tilde {\alpha}_{k}(0)  \right\rangle,j<k
\end{aligned}
\right. 
,
\label{modified_OTOC}
\end{equation}
where $\tilde{\alpha}_j(t)\equiv P^\dagger \alpha_j(t)$ with  $P = \prod_j \tau_j$ being the parity operator satisfying $P^3 =1$ and  $[H,P]=0$. Mathematically, we can prove that $F_{j,k}=\tilde{F}_{j,k}$ \cite{SuppAQSPA}. Now, the left-string operator $\alpha_{2k+1} = \left(\prod_{n<k} \tau_n\right) \sigma_{k+1}$  changes  into the right-string operator $\tilde{\alpha}_{2k+1}  = \left(\prod_{n \ge k} \tau^\dagger_n\right) \sigma_{k+1}$, which restores the distance between two operators in computing OTOCs via the MPO approach. Our algorithm is pictorially illustrated in Fig. \ref{Algorithm Illustration} with more details given in the Supplementary Materials  \cite{SuppAQSPA}.

\begin{figure}
  \centering\includegraphics[width=0.49\textwidth]{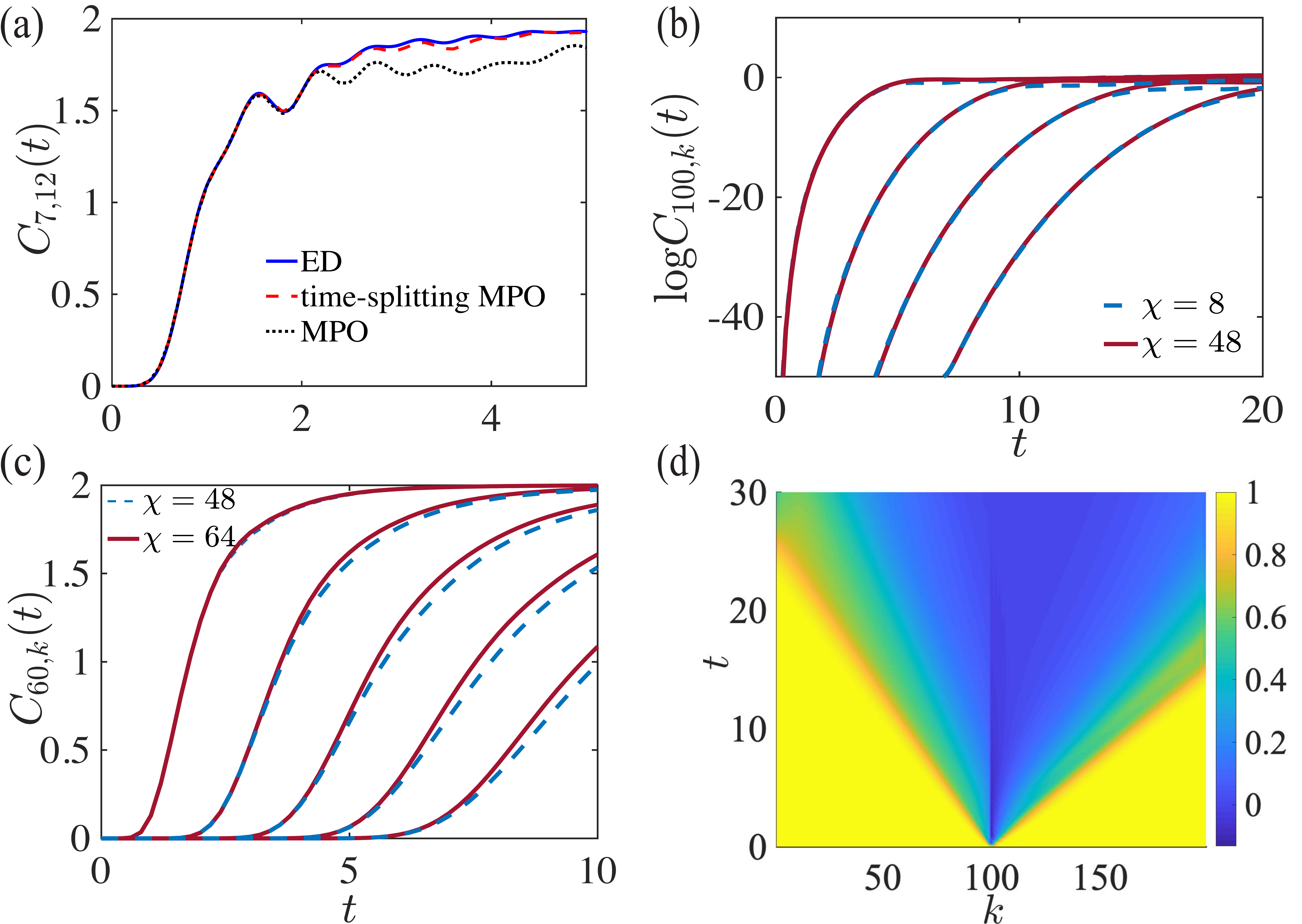}
  \caption{(a) A comparison between results from the MPO algorithms  and exact diagonalization (ED). Here, the bond dimension $\chi=48$ is used. (b) The  results of $C_{j,k}$ in the early-growth regime  by the MPO algorithm with  $j=100$ and  $k=80,60,40,20$.  (c) The results of $C_{j,k}$ for the  later-time regime with $j=60$ and $k=50,40,30,20,10$. (d) The light-cone structure of the OTOC  $\text{Re}[F_{j,k}(t)]$ is plotted with $j=100$.  We fix $\theta=\phi=0$ and other parameters are chosen as: $t_2=1$ for (b) and $t_2=0.5$ for (a), (c) and (d); $L=14, 200, 120,200$ for (a), (b), (c), and (d), respectively.   
  }
  \label{fig2}
\end{figure}

For the time-evolved MPOs in the early growth regime (before the wavefront reaches the left side of the right-string operator $\tilde{\alpha}_{k} $ ) , the truncation error is bounded owning to the entanglement lightcone structure and wiped off by the average over the infinite-temperature ensemble. To access the OTOC for a longer time, we may use the time-splitting MPO method: $\tilde{F}_{j,k}(t)=\tilde{F}_{j,k}^\text{late}(t) = \langle\alpha_{j}^{\dagger}(t/2) \tilde{\alpha}_{k}^{\dagger}(-t/2) \alpha_{j}(t/2) \tilde{\alpha}_{k}(-t/2)\rangle$ for $j<k$,  where we evolve both local parafermion operators in  forward and backward directions. With this method, we can capture the information scrambling in parafermion chains in both early-time and later-time regimes.  

{\it Light-cone structure.---}We now study the scrambling of information for parafermion chains. We first benchmark the effectiveness and accuracy of our algorithm. In  Fig.\ref{fig2}(a), we compare the MPO results with that from the exact diagonalization (ED) for a short parafermion chain with $L=14$. We find that with a moderate bond dimension ($\chi=48$), the MPO method without time-splitting works excellently for the entire early-growth regime, whereas for later times it becomes inaccurate due to the growth of entanglement. In contrast, the time-splitting MPO method works for both the early-time and later-time regimes with relative error smaller than $1\%$. In the following, we will use the the time-splitting MPO method with a small Trotter step $dt=0.002$ by default.

Then we compute the OTOCs for much longer parafermion chains, which are far beyond the capability of the ED method. In Fig.\ref{fig2}(b), we plot the result of $C_{j,k}$ in the early-growth regime with the system size $L=200$. It is clear that the curves for bond dimension $\chi=8$ match almost precisely with that for $\chi=48$, indicating that a small bond dimension is sufficient for computing OTOCs in the early-growth regime. For the later-growth regime, we also calculate $C_{j,k}$ with different bond dimensions for a parafermion chain with system size $L=120$, and our result is shown in Fig.\ref{fig2}(c). We find that the curves for $\chi=48$  match that for $\chi=64$ in the regime $C_{j,k}<0.4$, but after that deviations will show up owning to the growth of entanglement.

 
The above discuss have clearly demonstrated the effectiveness of our MPO method in computing OTOCs for parafermion chains in the entire early-growth regime. Truncation to small bond dimension only results in errors after the wavefront, and the scrambling of information ahead of and up to the wavefront can be captured accurately with our approach. Now, we discuss the anomalous quantum information scrambling for parafermion chains. First, we note that the model in Eq. (\ref{para_model}) is integrable when $\theta =\phi=0$ and $t_2=0$, where the OTOCs map out a symmetric light cone \cite{SuppAQSPA}, similar to the cases for conventional fermions or bosons. However, as shown in Fig. \ref{fig2}(d),  when we turn on the next-nearest-neighbor hoppings ($t_2\neq 0$) the light cone will become asymmetric, implying that the information propagation is asymmetric for the left and right directions. We stress that from the perspective of parafermions, the Hamiltonian is fully left-right symmetric when $\theta =\phi=0$.
The dynamic broken of the left-right symmetry is a reflection of anyonic statistics of the parafermions.


\begin{figure}
  \centering\includegraphics[width=0.49\textwidth]{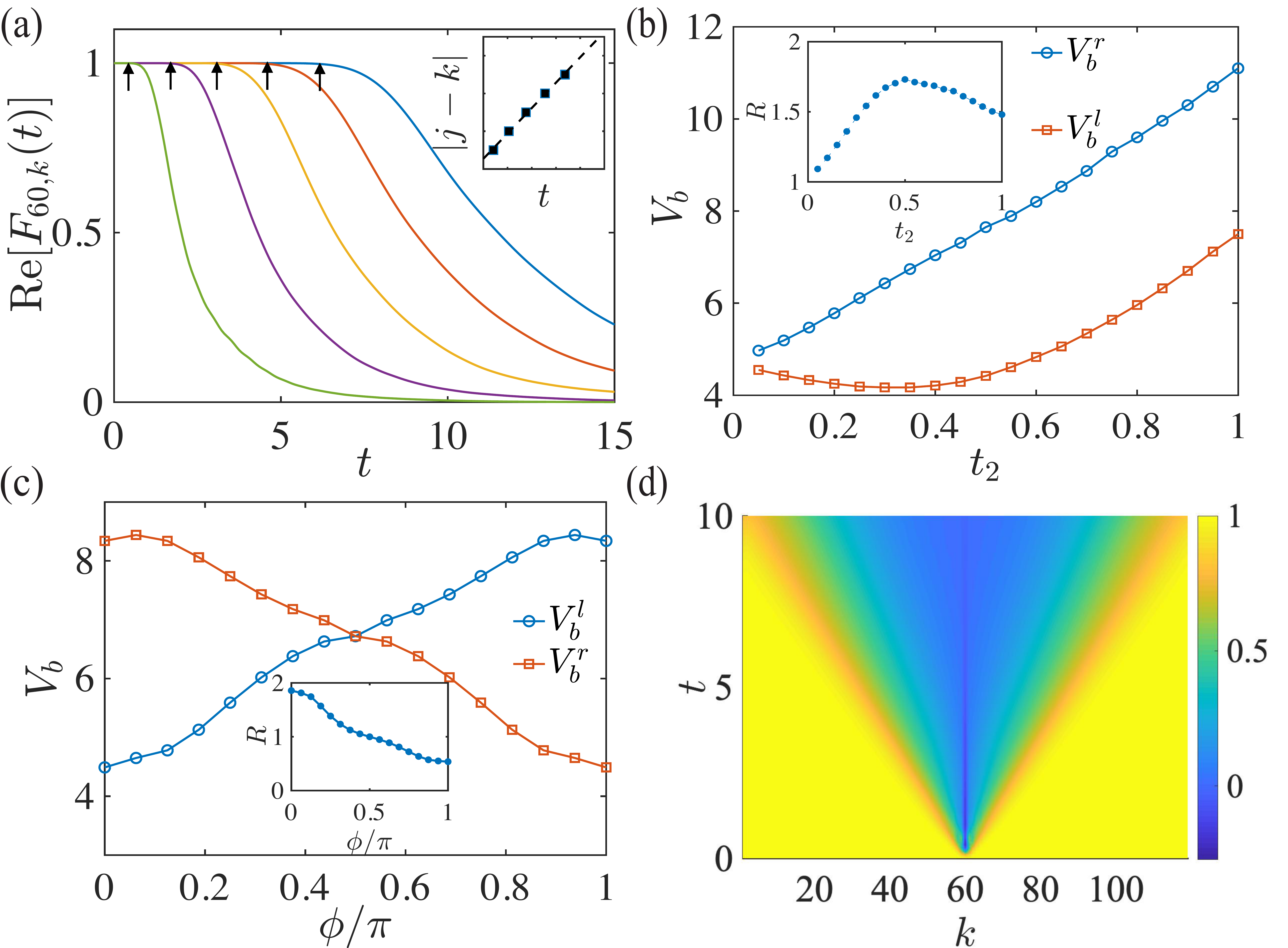}
  \caption{ (a) The dynamics of the OTOCs for a parafermion chain with length $L=120$, which characterizes the quantum information spreading from the middle of the chain ($j=60$) to the left $k=50,40,30,20,10$. The black arrows mark the positions where the OTOCs drop $1\%$ of their initial values, which are used to extract the butterfly velocity through linear fit as shown in the inset.  Here, the bond dimension $\chi=32$ is used and other parameters are chosen as $\theta=\phi=0$ and $t_2=0.9$. 
  (b) The extracted  left ($V_b^l$) and right ($V_b^r$)  butterfly velocities  as a function of $t_2$. The inset shows the ratio $R=V_b^r/V_b^l$ versus $t_2$.  (c) Dependence of $V_b^l$ and $V_b^r$ on $\phi$ when $\theta=\pi/6$ and $t_2=0.5$, with the inset showing the corresponding ratio. (d) The symmetric light-cone structure of the OTOCs  with $\theta=\pi/6$,  $\phi=\pi/2$, and $t_2=0.5$.  
  }
  \label{fig3}
\end{figure}

A more precise way to quantify the asymmetry of the information spreading is to utilize the butterfly velocity  $V_b^l$  ($V_b^r$) for the left (right) directions. We defined the butterfly velocity $V_b^{l/r}$ by the boundary of the space-time region where $\text{Re}(F_{j,k})$ drops by at least $1 \%$ of its initial values, as marked by arrows in Fig.\ref{fig3}(a). 
The linear fits of butterfly velocities $V_b^{l,r}$ with varying $t_2$ are shown in Fig. \ref{fig3}(b), from which it is clear that $V_b^r>V_b^l$ for the whole region $t_2>0$, indicating that information scrambles faster to the right direction. In addition, it is also interesting to note that  $V_b^r$ increases monotonically as  $t_2$ increases. Whereas, the dependence of $V_b^l$ on $t_2$ is non-monotonic: it decreases at first and then increases.  A maximum deviation of  $V_b^l$  from $V_b^r$  occurs around $t_2=0.5$.  In Fig.\ref{fig3}(c), we plot $V_b^{l,r}$ with varying $\phi$ and fixed $\theta=\pi/6$. Interestingly, $V_b^{l,r}$ have a crucial dependence on $\phi$: one can make information scrambles faster to the right (or left) direction by tuning  $\phi$. When $\phi=\pi/2$, we find that $V_b^l=V_b^r$ and the light cone is fully symmetric, as shown in Fig. .\ref{fig3}(d).


\begin{figure}
  \centering\includegraphics[width=0.49\textwidth]{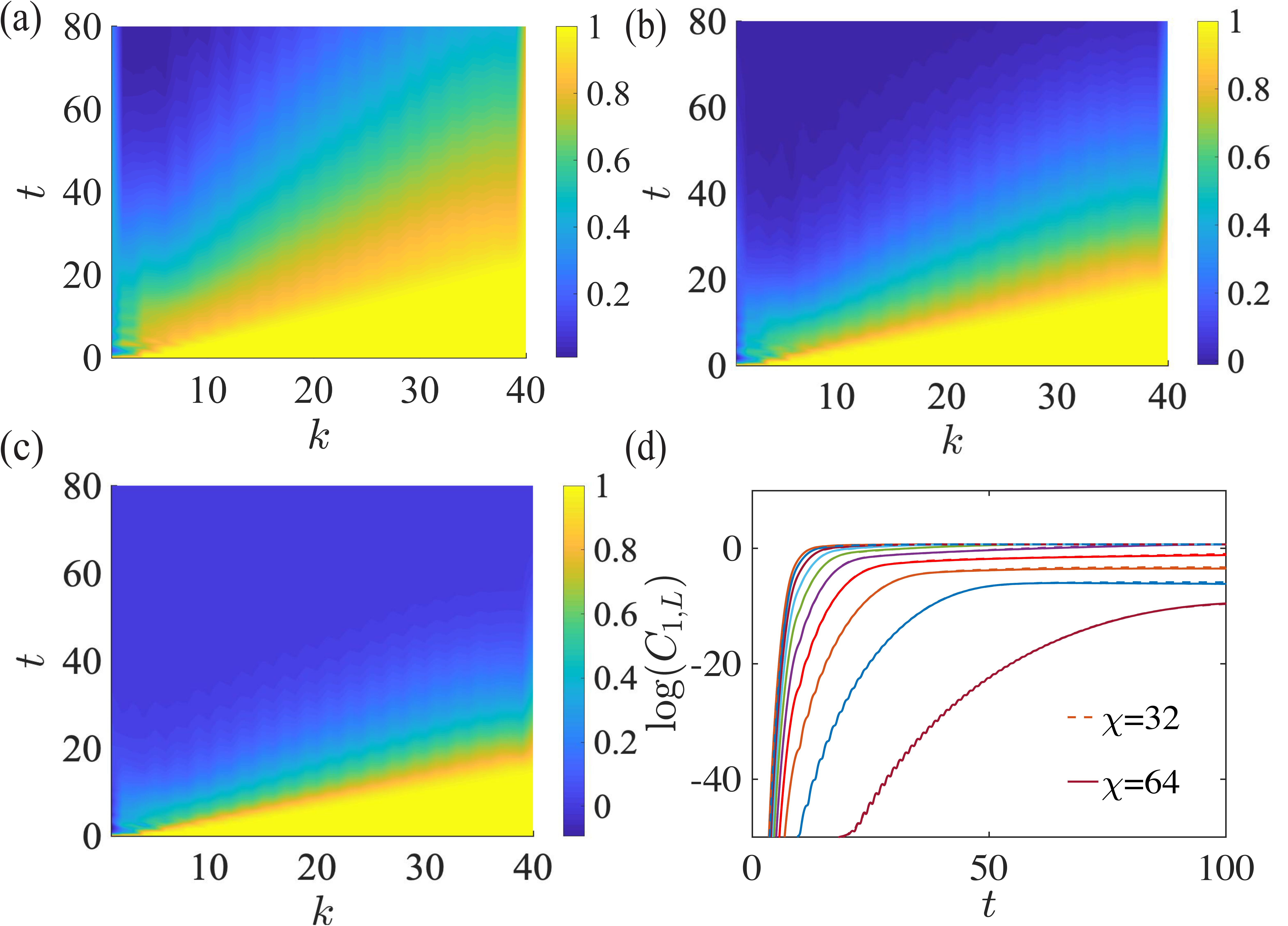}
  \caption{ The light-cone structures of the OTOCs  $\text{Re}[F_{1,k}(t)]$  for the parafermion chain model defined in Eq. (\ref{AlternatingPC}), with $\varphi=-\pi/6$, $J_1=1$, $L=40$, and $J_2=0.4,0.5,0.6$ respectively in (a), (b), and (c).   The bond dimension is $\chi=48$.  (d) The results of  $C_{1,L}$ with varying $J_2=0.1,0.2,\cdots,1.0$ from right to left. The curves for $\chi=32$ matches that of $\chi=64$ precisely, indicating  a negligible truncation error for the MPO algorithm in computing these OTOCs. }
  \label{fig4}
\end{figure}

{\it Symmetry analysis.---} In the Fig.\ref{fig3}(c), we find that $V^l_b(\phi)=V^r_b(\pi-\phi)$ and the preferred information scrambling direction can be reversed by sending $\phi \rightarrow \pi -\phi$. Here,  we show that this observation can be understood from the symmetry analysis of the Hamiltonian.  In fact, one can use two successive transformations, $(\sigma \rightarrow \sigma^\dagger, \tau \rightarrow \tau^\dagger)$ and $(\tau \rightarrow  \omega^{-1} \tau $, $\sigma_{2j} \rightarrow \omega^{-1} \sigma_{2j}$,  $\sigma_{2j+1} \rightarrow \sigma_{2j+1})$, to obtain $H_c (\theta=\pi/6, \phi) \rightarrow -H_c (\theta=\pi/6, \pi -\phi) $  and $\alpha_j^{\phi}(t) \rightarrow \alpha_j^{(\pi-\phi)\dagger}(-t) $, where the $\alpha_j^{\phi}(t)$ donates the evolution of a parafermion $\alpha_j(0)$ under the  Hamiltonian $H_c (\theta=\pi/6, \phi)$  \cite{SuppAQSPA}. Noting in addition that $\langle \alpha_j^\dagger(t) \alpha_k^\dagger(0)\alpha_j(t) \alpha_k(0)\rangle = \langle \alpha_j^\dagger(0) \alpha_k^\dagger(-t)\alpha_j(0) \alpha_k(-t)\rangle$, we thus obtain $C_{j,k}^{\phi}(t)=C_{k,j}^{\pi-\phi}(t)$, which explains the inversion symmetry between curves of $V^l_b$ and $V^r_b$. Particularly, when $\phi=\pi/2$ we have $C_{j,k}^{\pi/2}(t)=C_{k,j}^{\pi/2}(t)$, giving rise to the fully symmetric light cone shown in Fig.\ref{fig3}(d).

{\it Scrambling for strong zero modes.---}Strong zero modes lead to degeneracies across the entire spectrum and thus may offer potential advantages in building fault-tolerant qubits that works at high temperatures. With the introduced MPO algorithm, we are able to study information scrambling for strong zero modes at even infinite temperature. To this end, we consider the following parafermion chain model with alternating nearest-neighbour couplings \cite{jermyn2014stability,Fendley_2012}:
\begin{equation}
H = -J_1\sum_{j} e^{i \varphi}  \alpha_{2j}^\dagger\alpha_{2j+1}  -  J_2\sum_{j} \alpha_{2j-1}^\dagger\alpha_{2j} +  \text{H.c.}. \label{AlternatingPC}
\end{equation}
The stability of the zero modes in this model has been discussed  and the regime where the strong zero modes may exist has also been  estimated based on perturbation analysis and density matrix renormalization group algorithm near the ground states \cite{jermyn2014stability}. In the limit $J_2\rightarrow 0$,  the outermost parafermion operators drop out from the Hamiltonian (similar as in the Kitaev chain for Majoranas \cite{kitaev2001unpaired}) and represent localized zero modes that guarantee a threefold degeneracy for the whole spectrum. However, unlike the Majorana case for this $\mathbb{Z}_3$ parafermion chain there are strong evidences that localized zero modes disappear completely upon introducing arbitrarily small $J_2$ when $\varphi=0$, which is rather counterintuitive given that the system is in a gapped topological phase. Whereas, for nonzero $\varphi$ stable localized zero modes seems to survive small nonzero $J_2$ indeed \cite{jermyn2014stability,Fendley_2012}.

For our purpose, we compute the OTOC of two parafermion operators at the open ends $F_{1,L}(t)$ and our results are shown in Fig. \ref{fig4}. Here, we choose $\varphi=-\pi/6$ since at this point the zero modes are suspected to be the most robust \cite{jermyn2014stability,moran2017parafermionic}. In Fig.\ref{fig4}(a-c), we plot  $\text{Re}[F_{1,k}(t)]$  with  $L=40$, $J_1=1$ and varying $J_2$.  From Fig.\ref{fig4}(a), we see a sharp peak at the boundary of the light cone for $J_2=0.4$, indicating a drastically  lengthened scrambling time for the zero modes localized at the ends of the chain. Given that the our OTOC is calculated at infinite temperature, this sharp peak is a clear-cut evidence of the existence of strong zero modes at the ends of the parafermion chain for nonzero $J_2$. When $J_2$ increases, this peak diminishes and nearly disappears when $J_2=0.6$, as shown in Fig.  \ref{fig4}(b,c). This implies the transition point is in the regime $J_2=0.4\sim 0.6$, which is consistent with the perturbative analysis in Ref. \cite{jermyn2014stability}. To see it more clearly,  we calculate the $C_{1,40}$ for increasing $J_2$ in Fig.\ref{fig4}(d).  As we can see,  for a fixed time window $t<100$, the squared commutator $C_{1,L}$ increases rapidly and saturate to its maximum value very soon for $J_2\ge0.6$.  While for $J_2<0.6$ , it increases much slower, indicating the presence of strong zero modes as well. 

{\it Discussion and conclusion.---} A number of protocols for measuring OTOCs in various systems have been proposed \cite{swingle2016measuring,zhu2016measurement,yao2016interferometric,halpern2017jarzynski,halpern2018quasiprobability,
campisi2017thermodynamics,yoshida2017efficient}. Indeed, recently experimental measurement of OTOCs has been demonstrated with trapped ions \cite{garttner2017measuring} and nuclear magnetic resonance quantum simulators \cite{Wei2018Exploring,Li2017Measuring}. For $\mathbb{Z}_3$ parafermions, different blueprints for their experimental realization have also been introduced in a variety of systems, ranging from lattice defects in fractional Chern insulators \cite{vaezi2014superconducting} and fractionalized topological insulators/superconductors \cite{Klinovaja2014Kramers,cheng2012superconducting} to quantum Hall bilayers \cite{barkeshli2014synthetic,peterson2015abelian} and bosonic cold atoms \cite{maghrebi2015parafermionic}. Yet, to the best of our knowledge, no experimental proposal of measuring OTOCs for $\mathbb{Z}_3$ parafermions has been introduced hitherto. In the future, it would be interesting to study how OTOCs for parafermion chains can be measured in experiment and consequently observe the anomalous information scrambling predicted in this paper. 

In summary, we have introduced a low-cost MPO algorithm to calculate the OTOCs for parafermion chains, which can capture the scrambling of quantum information in the entire early-growth regime with modest bond dimension. With this powerful algorithm, we have explored the anomalous information dynamics for $\mathbb{Z}_3$ parafermion chains up to a system size far beyond the capability of previous numerical approaches. We found that information can scramble both symmetrically and asymmetrically for parafermion chains, even for inversion-invariant Hamiltonians involving merely hopping terms. In addition, we found a deformed light cone structure with a sharp peak at the boundary, which offers a unambiguous evidence of the strong zero modes  at infinite temperature.  Although we have only focused on $\mathbb{Z}_3$ parafermions, our introduced algorithm applies to the general $\mathbb{Z}_n$ parafermions and Abelian anyons (such as the anyon-Hubbard model) as well. Our results  not only provide a powerful method for accessing quantum information scrambling in systems with exotic quasiparticles, but also uncover the peculiar information dynamics  for parafermions which would benefit future studies in both theory and experiment.


We acknowledge helpful discussions with  Fang-Li Liu and Sheng-Long Xu. This work is supported by the start-up fund from Tsinghua University (Grant. No. 53330300320), the National Natural Science Foundation of China (Grant. No. 12075128), and the Shanghai Qi Zhi Institute. 


\bibliography{ref}

\clearpage
\setcounter{figure}{0}
\makeatletter
\renewcommand{\thefigure}{S\@arabic\c@figure}
\setcounter{equation}{0} \makeatletter
\renewcommand{\thesection}{S.\Roman{section}}
\renewcommand \theequation{S\@arabic\c@equation}
\renewcommand \thetable{S\@arabic\c@table}

\begin{center} 
{\large \bf Anomalous Quantum Information Scrambling for $\mathbb{Z}_3$ Parafermion Chains}
\end{center}

\section{The MPO algorithm } \label{sec:algorithm}
In the main text, we have given a brief introduction to the MPO algorithm for calculating the OTOCs in parafermion chains. Here we generalize this method to some other models which consist of  on-site symmetries and give  more details of the MPO algorithm.

\subsection{General models}
Our algorithm is generally applicable for Hamiltonians which obey certain  on-site symmetry,  that is
\begin{equation}
S= \otimes_n u_n.
\end{equation}
Here, $u_n$ is an on-site operator on the $n$ site and $[S,H]=0$.  A  general OTOC can be written as
\begin{equation}
F_{ij} = \langle W_i^\dagger(t) V_j^\dagger W_i(t) V_j  \rangle,
\end{equation}
where $W,V$ are unitary operators. By inserting the identity operator $\mathbb{I}=SS^\dagger$ into the OTOC,  we obtain
\begin{equation}
F_{ij} =  \langle  e^{iHt} W_i^\dagger e^{iHt}  V_j^\dagger S S^\dagger e^{iHt} S S^\dagger W_i  S S^\dagger e^{-iHt} S S^\dagger V_j  \rangle.
\label{eq:transform}
\end{equation}
If we add the condition $S^\dagger W_i  S \sim W_i $ ($\sim$ means equal up to a constant),  accompanying with $S^\dagger e^{iHt} S = e^{iHt}$,  the OTOC $F_{ij}$ reduces to
\begin{equation}
F_{ij}  \sim  \langle  W_i^\dagger(t)  (V_j^\dagger S) W_i(t) (S^\dagger V_j)  \rangle.
\end{equation}
Then the OTOC between $W_i$ and $V_j$ is equivalent to the OTOC between $W_i$ and $S^\dagger V_j$.

Taking anyon-Hubbard model for example.  The anyon-Hubbard model can be written as
\begin{equation}
H_A= -J\sum_{j=1} \( a_j^{\dagger} a_{j+1} + \Hc \) + \frac{U}{2} \sum_{j=1} n_j (n_j-1),
\label{ahm}
\end{equation} 
and the OTOC is defined as 
\begin{equation}
F_{jk}^A(t)= \Braket{a_j^{\dagger}(t) a_k^{\dagger}(0) a^{\phantom\dagger}_j(t) a^{\phantom\dagger}_k(0) } e^{i\theta \sgn(j-k)}.
\end{equation}
The Hamiltonian has symmetry $S_A = e^{-i\theta\sum_j a_j^{\dagger} a_{j}}$. By utilizing the same transformation as Eq. \ref{eq:transform} ,  $F_{jk}^A(t)$ is transformed to
\begin{equation}
F_{jk}^A(t)= \Braket{a_j^{\dagger}(t) \left(a_k^{\dagger}(0)S_A\right) a^{\phantom\dagger}_j(t) \left(S_A^\dagger a^{\phantom\dagger}_k(0)\right) } .
\end{equation} 
Since  $a_k = b_k  e^{-i\theta\sum_{j<k} a_j^{\dagger} a_{j}} $ is a left string operator,   $\left(S_A^\dagger a_k(0)\right)$ become a right string operator.  Here,  $ b_k$
 is the Boson annihilation operator. 
  
 In the next part,  we calculate the OTOCs in parafermion chains with $L=20$  as an illustrating example. 

\subsection{The details of the MPO algorithm }
In the main text, we have introduced the algorithm from entanglement points of view,  especially emphasized the importance of the distance between two local operators in the calculation of OTOCs.  In this section,  we give more details about the reason why this condition is essential for the MPO method to work well in the early-growth regime.  However,  this condition is not sufficient,  as we would mention below,  the calculation of OTOC with infinite-temperature ensembles is also important.

\begin{figure}
  \centering\includegraphics[width=0.5\textwidth, height=8cm]{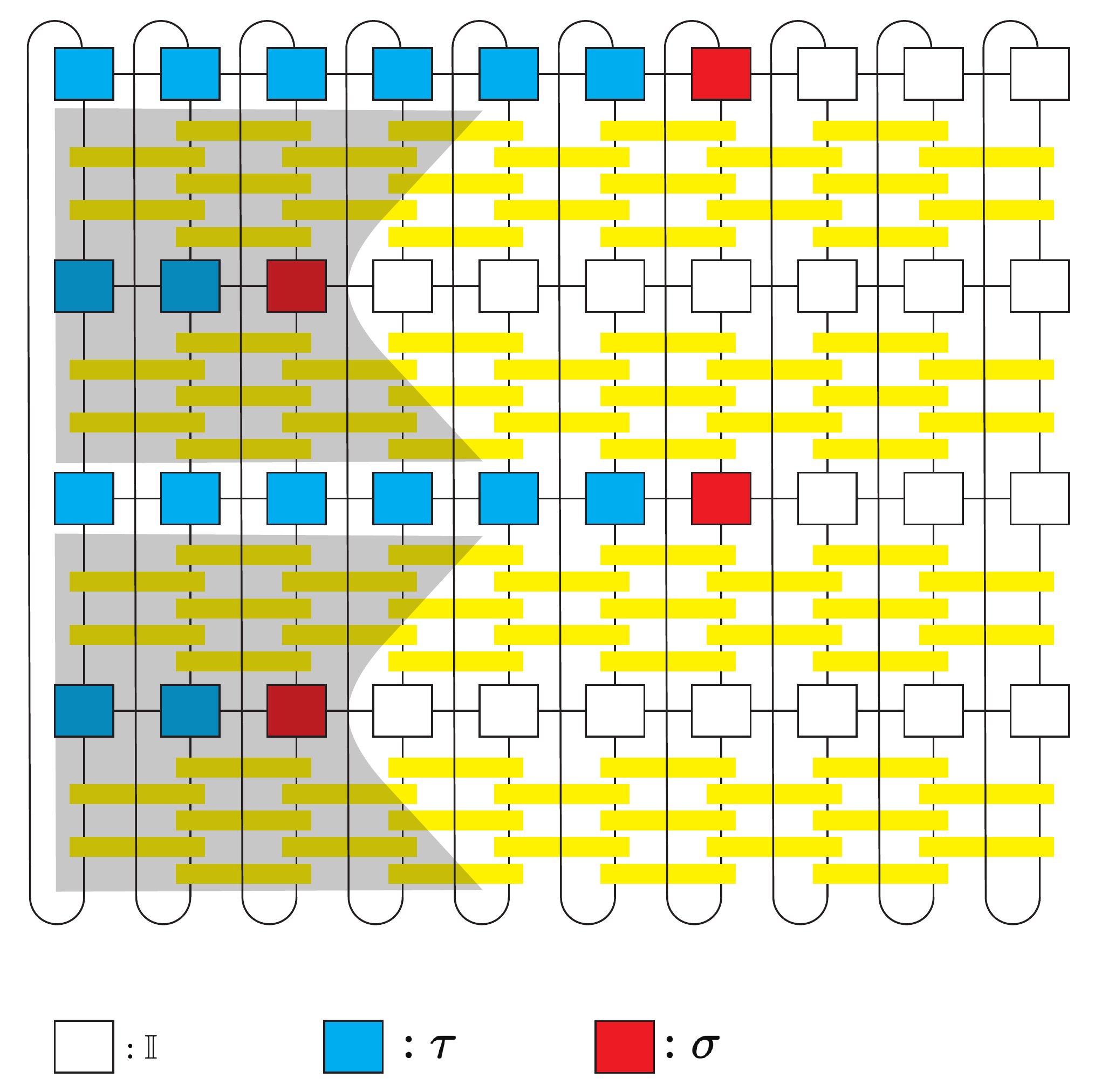}
  \caption{Graphical representation for the OTOC $F_{j,k} $($j<k$) in the natural tensor network form,  in contrast to the modified $\tilde{F}_{j,k} $ in Fig.1 in the main text.  Here,  we choose the parameter $j=5,k=13$.   
  }
  \label{figS1}
\end{figure}

Without loss of generality,  we consider the calculation of the OTOCs 
\begin{equation}
 F_{j,k}(t) = \left\langle\alpha_{j}^{\dagger}(t) \alpha_{k}^{\dagger}(0) \alpha_{j}(t) \alpha_{k}(0)\right\rangle\omega^{ \operatorname{sgn}(j-k)},
\end{equation}
in a  20-sites parafermion chain,  which can be mapped to a 10-sites $\mathbb{Z}_3$ spin chain.   Next,  we take $j=5,k=13$ for example.  The local parafermion operators can be written as
\begin{equation}
\begin{aligned}
\alpha_5 &= \tau_1 \tau_2 \sigma_3, \\
\alpha_{13} &= \tau_1 \tau_2  \tau_3 \tau_4 \tau_5 \tau_6 \sigma_7,
\end{aligned}
\end{equation}
and the general MPO calculation of  $F_{j,k}(t)$ is illustrated in Fig.\ref{figS1}.  We use blue(red) blocks to represent the operator $\tau$($\sigma$) and use white blocks to represent the identity operator. The MPO evolution is based on time-evolving block decimation (TEBD) method \cite{vidal2004efficient},  which is represented by yellow blocks.  The trace operation corresponds to contracting all the physical indices from top to bottom. The grey regions correspond to the induced operator light-cone expansion. However,  this direct MPO calculation could only capture the early time regime of OTOC growth,  due to the rapid growth of entanglement which induces huge  truncation errors.

In order to get rid of these truncation errors,  we calculate the modified but mathematically equivalent quantity obtained by inserting the parity operator $P=\prod_j \tau_j$ inside the OTOC:
\begin{equation}
\begin{aligned}
F_{5,13}(t) &= \left\langle\alpha_{5}^{\dagger}(t) \alpha_{13}^{\dagger}(0) P P^\dagger \alpha_{5}(t) P P^\dagger  \alpha_{13}(0)\right\rangle\omega^{ 2}\\
&= \left\langle\alpha_{5}^{\dagger}(t) \left( \alpha_{13}^{\dagger}(0) P \right)  \alpha_{5}(t)  \left(P^\dagger  \alpha_{13}(0) \right)\right\rangle\\
&= \left\langle\alpha_{5}^{\dagger}(t) \tilde{\alpha}_{13}^{\dagger}(0) \alpha_{5}(t)  \tilde{ \alpha}_{13}(0)\right\rangle \equiv \tilde{F}_{5,13}(t) ,
\end{aligned}
\end{equation}
where the operator  $\tilde{ \alpha}_{13} =\tau_7 \sigma_7 \tau_8 \tau_9\tau_{10} $ becomes right string form,   as  illustrated in Fig. 1 in the main text.

\begin{figure}
  \centering\includegraphics[width=0.5\textwidth, height=7.5cm]{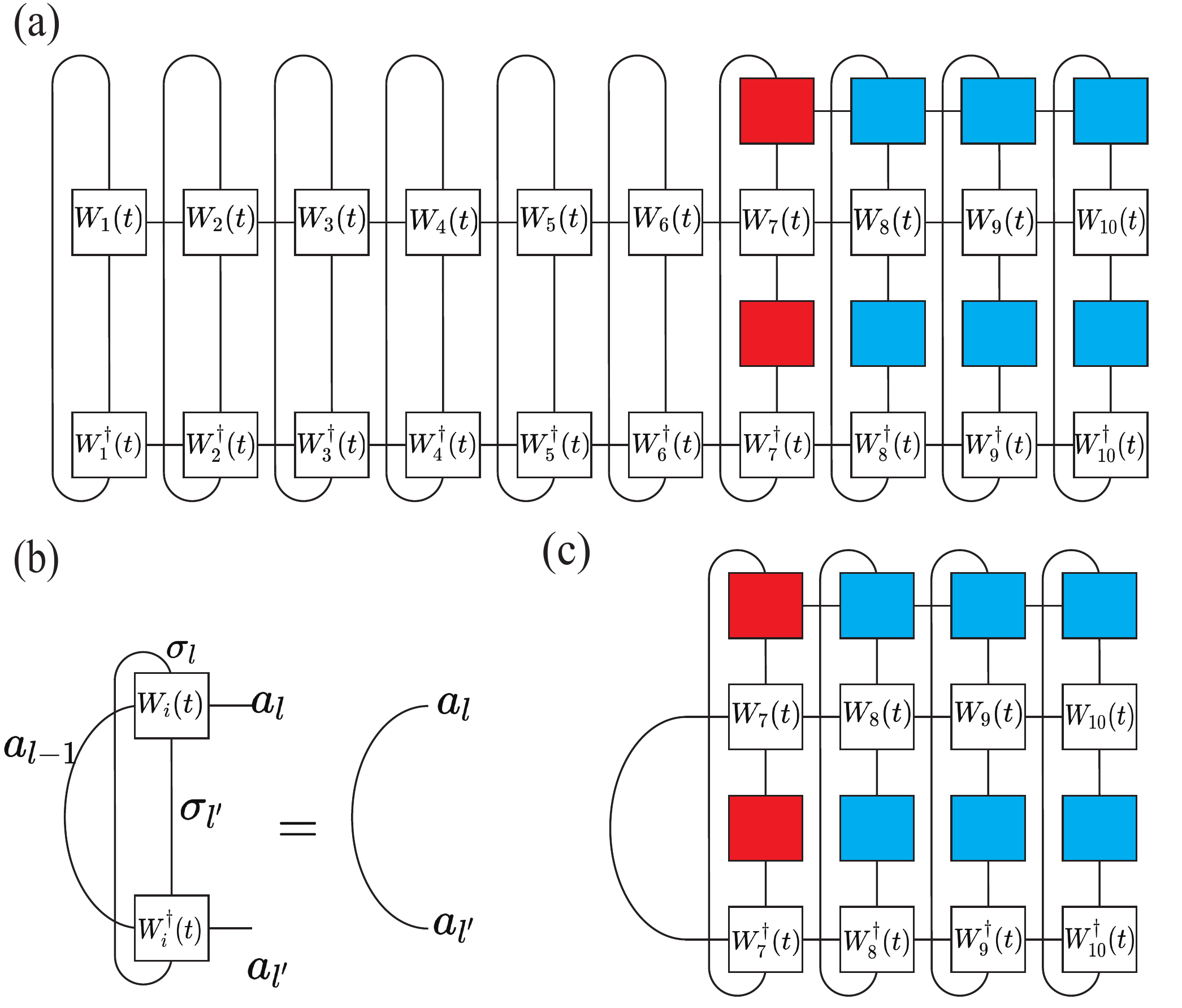}
  \caption{  (a) Reduction of the tensor network from the Fig.1 in the main text.  The $\alpha_5(t)$ is replaced by MPO $\mathrm{W}(t)$. (b) Left canonical condition of MPO tensors is illustrated in diagram,  $\sigma_l$ and  $\sigma_{l^\prime}$ are physical indices and $a_l, a_{l^\prime}, a_{l-1}$ are vurtual indices.  (c) Simplified MPO representation of  OTOC.
  }
  \label{figS2}
\end{figure}

The MPO calculation of $ \tilde{F}_{5,13}(t)$ could be simplified using the product form of $\tilde{ \alpha}_{13}$,  see Fig. \ref{figS2}(a).   Here,  we have replaced $\alpha_{5}^{\dagger}(t)$  with $\mathrm{W}(t)$ in MPO form,  and $W_i(t)$ are tensors of $W(t)$ at site $i$.   Then,  taking advantage of the left canonical condition of MPO
\begin{equation}
\sum_{\sigma_l ,\sigma_{l^\prime}} W_i^{\sigma_l ,\sigma_{l^\prime}\dagger} W_i^{\sigma_l ,\sigma_{l^\prime}} = \mathbb{I},
\end{equation}
which is shown in graphical representation in Fig.\ref{figS2}(b),  the MPO of OTOC  reduces to the structure of Fig.\ref{figS2}(c).  

Now we see that  the calculation of the OTOC is only related to the  contraction of the tensor right to the site $k$.  Therefore ,  as long as the truncation error is small for these local MPO tensors,  the MPO algorithm is effective.   In fact,  as the truncation error is confined by the light-cone in the early-growth regime,  our method is efficient.  It is worth mentioning that the left canonical condition entails the trace operation,  which means that the OTOC computed is   averaged at infinite-temperature.

\section{Hamiltonian  and dynamical symmetries } \label{sec:symmetry}
In the main text,  we have mentioned the relevant symmetries for the OTOC dynamics.  Here,  we give the detailed derivation for them.  

\subsection{Hamiltonian symmetry }
To  discuss the symmetry in an explicit way, we give the  representation of $\sigma_j$ and $\tau_j$ in the matrix form:
\begin{equation}
\sigma=\left(\begin{array}{lll}
0 & 1 & 0 \\
0 & 0 & 1 \\
1 & 0 & 0
\end{array}\right) , \quad \tau=\left(\begin{array}{ccc}
1 & 0 & 0 \\
0 & \omega & 0 \\
0 & 0 & \omega^{2}
\end{array}\right) , \quad \omega=e^{2 \pi i / 3}
\end{equation} 
First, we consider the symmetry of the mapped clock Hamiltonian and define the  inversion transformation $\mathcal{I}$ as:
\begin{equation}
\mathcal{I}: \sigma_j \rightarrow \sigma_{N-j+1}, \  \tau_j \rightarrow \tau_{N-j+1}.
\end{equation}
Then, we introduce the following time-reversal anti-unitary transformation $\mathcal{T}$ as
\begin{equation}
\mathcal{T}: \sigma_j \rightarrow \sigma_{j}, \  \tau_j \rightarrow \tau^\dagger_j.
\end{equation} 
The composition $\mathcal{T} \circ \mathcal{I}$ defines a new operation which leave the Hamiltonian invariant. This transformation changes the Hamiltonian $H_c$ as 
\begin{equation}
\begin{aligned}
\tilde{H}_c =& -t_1 \sum_{j=1}^{N-1} e^{-i\theta}\omega^{2} \sigma_{N+1-j}^\dagger \sigma_{N-j} - t_1 \sum_{j=1}^{N} e^{-i\theta}\omega^{2} \tau^\dagger_{N+1-j} \qquad \\ 
&+ t_2 \sum_{j=1}^{N-1}  e^{-i\phi}\sigma_{N+1-j}^\dagger \tau^\dagger_{N-j} \sigma_{N-j}  \\
&+ t_2 \sum_{j=1}^{N-1}  e^{-i\phi}\sigma_{N+1-j}^\dagger \tau^\dagger_{N+1-j} \sigma_{N-j} +  \Hc
\end{aligned}
\label{clock_model}
\end{equation}
From the parafermion perspective,   this transformation $\mathcal{T} \circ \mathcal{I}$ inverses the parafermion sites as 
\begin{equation}
\mathcal{T} \circ \mathcal{I}: \alpha_j \rightarrow \tilde{\alpha}_{-j}  \equiv P \alpha_{-j} ,
\end{equation}
and the  anti-unitary operation  preserve the commutation relation $\alpha_i \alpha_j = \alpha_j \alpha_i \omega^{\text{sgn}(j-i)} $.  The detailed transformation of $\alpha_j$ by $\mathcal{T} \circ \mathcal{I}$ is:
\begin{eqnarray}
\alpha_{2j-1} &=& \left(\prod_{k=1}^{j-1} \tau_k\right) \sigma_j \\
& \rightarrow & \tilde{\alpha}_{2N+2-2j} = \left(\prod_{k=N}^{N+2-j} \tau^\dagger_k\right) \sigma_{N+1-j} ,\\ 
\alpha_{2j}  &=& \omega^2 \left(\prod_{k=1}^{j-1}\tau_k\right)  \sigma_j \\
&\rightarrow  & \tilde{\alpha}_{2N+1-2j}= \omega \left(\prod_{k=N}^{N+1-j} \tau^\dagger_k\right)  \sigma_{N+1-j},
\end{eqnarray}
which leaves the commutation relation invariant
\begin{equation}
\tilde{\alpha}_j^3 =1,\  \tilde{\alpha}_j^\dagger =\tilde{ \alpha}_j^2, \  \tilde{\alpha}_i \tilde{\alpha}_j = \tilde{\alpha}_j \tilde{\alpha}_i\omega^{\text{sgn}(j-i)}.
\end{equation}
However, this symmetry does not  guarantee the symmetry of the OTOC dynamics.

\subsection{Dynamical symmetry }
In the main text,  the special line $\theta=\pi/6, \phi=\pi/2$  exhibits  OTOC dynamical symmetry with regard to the parafermion chain model. We consider the effect of  two successive transformations on the OTOC. Initially, we assume the Hamiltonian is $H_c (\theta=\pi/6, \phi)$. We define a transformation $\mathcal{U}$ as
\begin{equation}
\mathcal{U}:\sigma \rightarrow \sigma^\dagger, \tau \rightarrow \tau^\dagger,
\end{equation}
where $\mathcal{U}$ is a unitary transformation with matrix representation 
\begin{equation}
U=\left(\begin{array}{lll}
1 & 0 & 0 \\
0 & 0 & 1 \\
0 & 1 & 0
\end{array}\right),
\end{equation}
and the transformation maps the Hamiltonian as
\begin{equation}
H_c (\theta=\pi/6, \phi) \rightarrow H_c (\theta=-\pi/6, 2\pi/3-\phi) .
\end{equation}
In the second transformation, we redefine
\begin{equation}
\tau \rightarrow  \omega^{-1} \tau , \quad \sigma_{2j} \rightarrow \omega^{-1} \sigma_{2j},  \quad \sigma_{2j+1} \rightarrow \sigma_{2j+1},
\end{equation}
the Hamiltonian changes as
\begin{equation}
H_c (\theta=-\pi/6, 2\pi/3-\phi) \rightarrow -H_c (\theta=\pi/6, \pi-\phi).
\end{equation}
Then under the two successive transformations, the Hamiltonian changes as 
\begin{equation}
H_c (\theta=\pi/6, \phi) \rightarrow -H_c (\theta=\pi/6, \pi-\phi).
\end{equation}
Whereas the parafermion  operator changes as 
\begin{equation}
\alpha_j^{\phi}(t) \rightarrow \alpha_j^{(\pi-\phi)\dagger}(-t),
\end{equation}
where  $\alpha_j^{\phi}(t)$ is defined by
\begin{equation}
\alpha_j^{\phi}(t)\equiv e^{iH_c(\theta=\pi/6,\phi)} \alpha_j e^{-iH_c(\theta=\pi/6,\phi)} .
\end{equation}
Then  $ F_{j,k}=\langle \alpha_j^{\phi\dagger}(t) \alpha_k^\dagger(0)\alpha^\phi_j(t) \alpha_k(0)\rangle \omega^{ \operatorname{sgn}(j-k)} $ changes to
\begin{equation}
\langle \alpha_j^{\pi-\phi}(-t) \alpha_k(0)\alpha_j^{(\pi-\phi)\dagger}(-t) \alpha_k^\dagger(0)\rangle \omega^{ \operatorname{sgn}(j-k)}.
\end{equation}
Using the identity
\begin{equation}
\langle \alpha_j^\dagger(t) \alpha_k^\dagger(0)\alpha_j(t) \alpha_k(0)\rangle = \langle \alpha_j^\dagger(0) \alpha_k^\dagger(-t)\alpha_j(0) \alpha_k(-t)\rangle,
\end{equation}
 as well as the property of trace operation,  the OTOC  becomes
\begin{equation}
 \langle \alpha_j^\dagger(0) \alpha_k^{(\pi-\phi)\dagger}(t)\alpha_j(0) \alpha^{\pi-\phi}_k(t)\rangle \omega^{ \operatorname{sgn}(j-k)} \equiv F_{k,j}^\dagger .
\end{equation}
Finally,  taking advantage of $C_{i,j}=2(1-\text{Re}(F_{i,j}))$,   we obtain
\begin{equation}
C^{\phi}_{j,k}(t) \rightarrow C^{\pi-\phi}_{k,j}(t),
\end{equation}
which explains the symmetry of dynamics in the main text.

\section{Level statistics } \label{sec:level}

\begin{figure}
  \centering\includegraphics[width=0.5\textwidth, height=14cm]{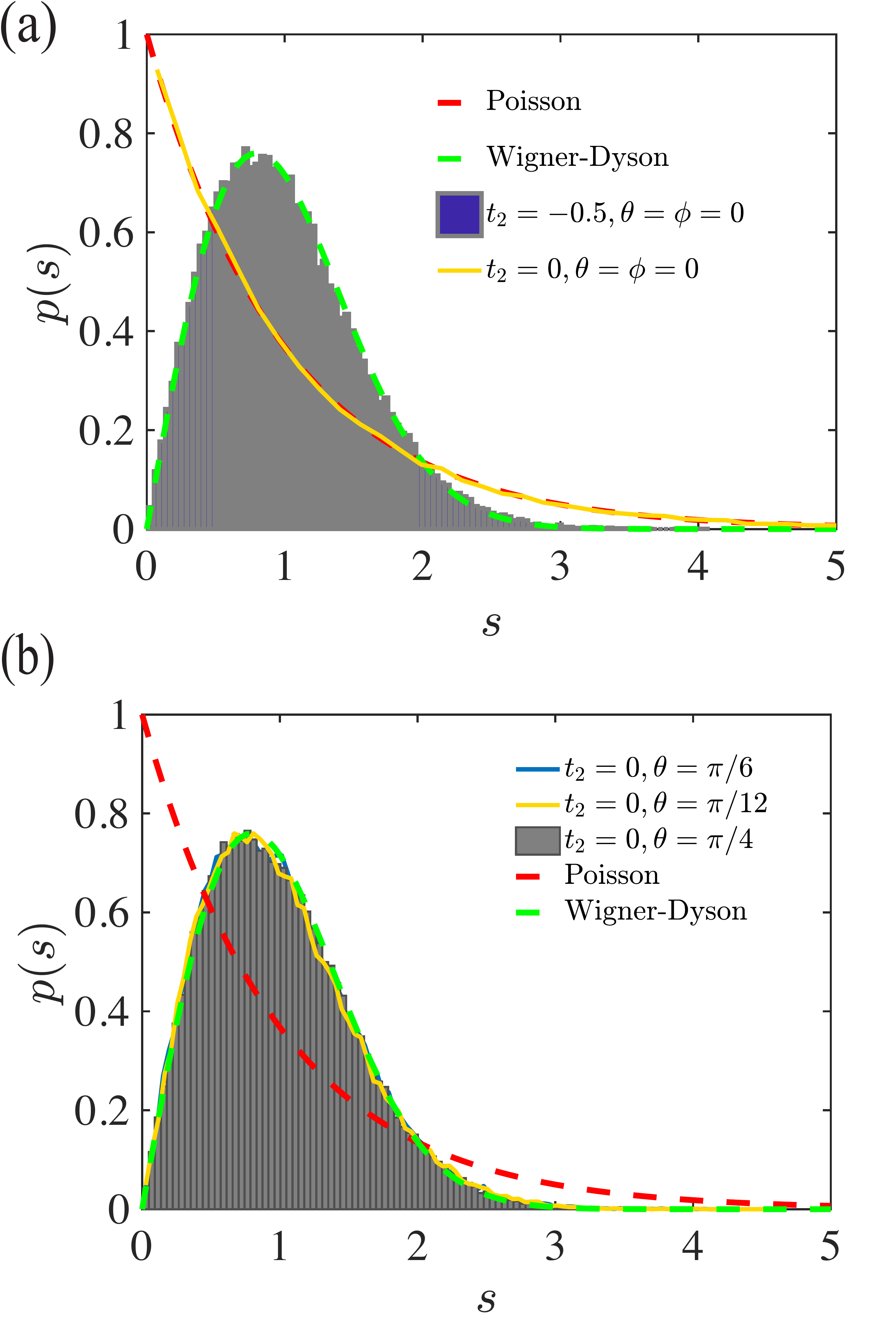}
  \caption{ Level space distribution of the spectra for model Eq.1 in the main text with chain length $L=12$.
  The spectra are restricted in parity $P=0$ subspace and the results for $P=1$ and $P=2$ are similar.   Dashed curves are $P(s)=e^{-s}$(red) and $P(s)=\frac{\pi}{2} s \mathrm{e}^{-\frac{\pi}{4} s^{2}}$(green),  typical for integrable or quantum chaotic systems.
  }
  \label{figS3}
\end{figure}

To  check whether a many-body Hamiltonian in a certain regime is  integrable or quantum chaotic,  we calculate the level space distribution of the spectra, which is a strong indicator for quantum chaos \cite{haake1991quantum}.

In Fig.\ref{figS3}(a-b),  we show the level space distribution for the Hamiltonian in Eq. (1) in the main text for different parameter regimes.   In the special point, where the next-nearest-neighbor interaction is turned off and $\theta=0$,  the model is integrable,  which is illustated in Fig. \ref{figS3}(a).  The levels show no repulsion and the probability distribution of spacings is approximately given by $P(s)=e^{-s}$.   Despite this point,  as we increase the next-nearest-neighbor interaction, see Fig. \ref{figS3}(a),  or add a no-zero chiral phase, see Fig. \ref{figS3}(b),  we  find a level repulsion and the statistics follows the Wigner-Dyson distribution.   The level spacing distribution has the following shape, $P(s)=\frac{\pi}{2} s \mathrm{e}^{-\frac{\pi}{4} s^{2}}$,  which indicates the nonintegrablility of the model.

\section{more numerical results } \label{sec:more}
\begin{figure}
  \centering\includegraphics[width=0.5\textwidth, height=6cm]{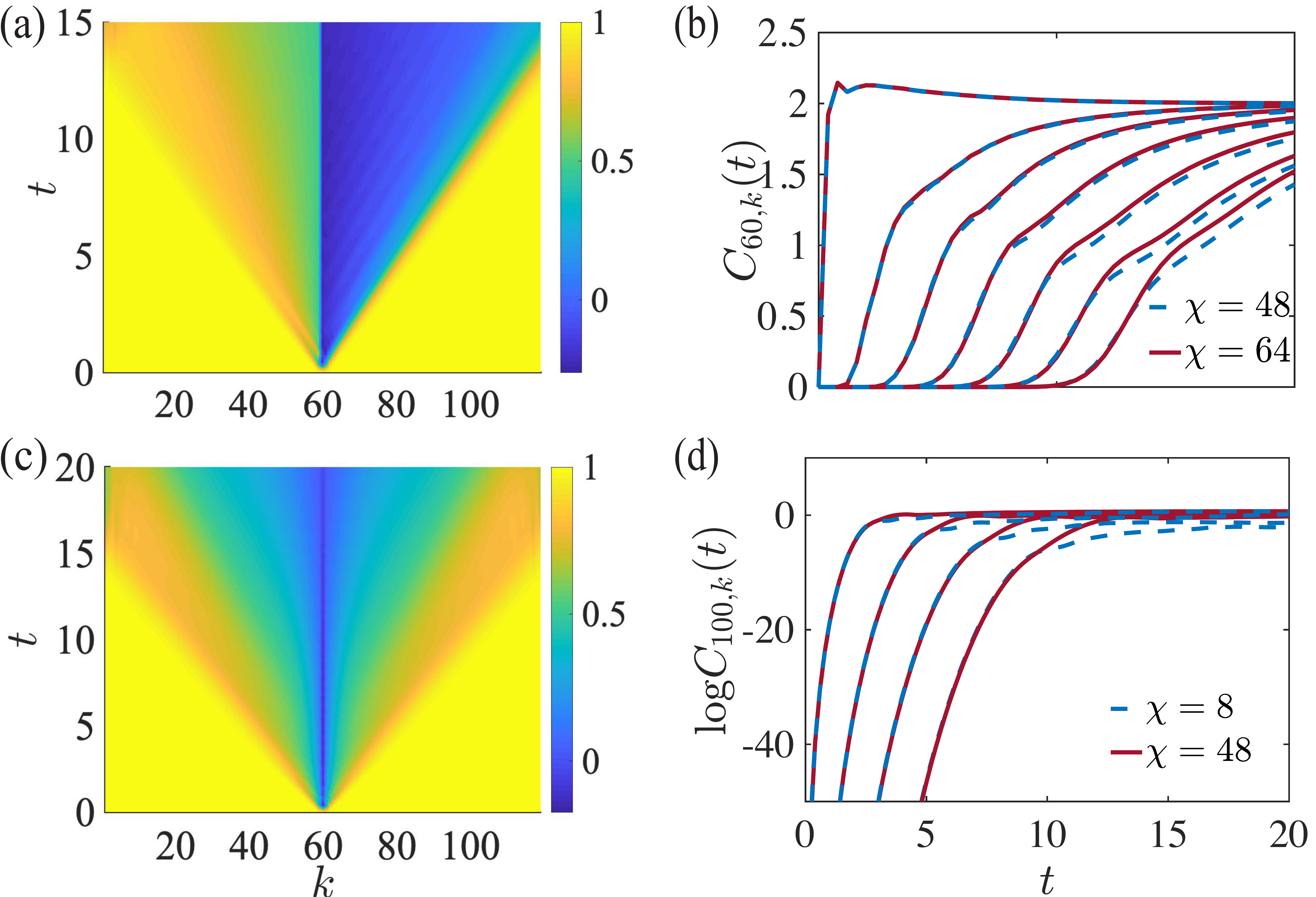}
  \caption{ OTOC growth $\text{Re}(F_{j,k})$ are obtained within MPO approach for differen  angles $\theta$,  $\phi$  in (a),(c)  and squared commutators $C_{j,k}$ are plotted in late and early-growth regime in (b),(d).  The light-cone structure of OTOC is ploted in (a) with parameters $\theta=\phi =0$, $t_2=0$ and $\theta=\pi/6$,  $\phi=\pi/2$, $t_2=0$ in  (c).  In both figures,  the color maps are interpolated to better illustrate the light-cone structure.    The  late-growth regime of  the squared commutator $C_{j,k}$ are ploted in (b) with parameters $\theta=0$, $\phi=0$, $t_2=0.5$,$j=60$ for varying $k=60,70,80,90,100,110$  and the  early-growth regime of  the squared commutator $C_{j,k}$  are plotted in (d)  with parameters $\theta=0$, $\phi=0$, $t_2=1$,$j=100$ for varying $k=120,140,160,180$. 
  }
  \label{figS4}
\end{figure}

In this section, we give more numerical results on the OTOC calculation.  We first consider the special case, when the angle $\theta=\phi =0$. We can clearly see from Fig. \ref{figS4}(a) that the information spreading asymmetric between two directions when the next-nearest-neighbor coupling is turned off. The information scrambles much faster to the right than to the left and it seems that there does not exist a clear wavefront  in the left-hand side.  This is due to the  integrability of the model with only nearest-neighbor couplings at the point $\theta=\phi=0$.  In Fig. \ref{figS4}(c),  we plot the OTOC in time-space with parameter  $\theta=\pi/6$,  $\phi=\pi/2$, $t_2=0$,  and find the light-cone structure is indeed symmetric, which is consistent with the symmetry analysis results in \ref{sec:symmetry}.  In Fig. \ref{figS4}(b), (d),  we calculate the OTOC both in the early and later growth regime and set $j<k$, which are in parallel with the results $j>k$ in the Fig. 2 in the main text.   These results indicate that the scrambling can be well captured by the MPO algorithm in both directions with modest bond dimension.

\section{The OTOC in the topological regime} \label{sec:decomposition}
The OTOC between parafermions at the two open ends can be written as
\begin{equation}
F_{1,L}=\frac{1}{3^{L/2}} \operatorname{Tr}\left(\alpha_{1}^{\dagger}(t) \alpha_{L}^{\dagger} \alpha_{1}(t) \alpha_{L}\right) \omega^{-1}.
\end{equation}
By utilizing the energy eigenstates as a basis, $F_{1,L}$ can be expanded as:
\begin{eqnarray}\label{eq:decom}
F_{1,L}&=&\frac{\omega^{-1}}{3^{L/2}}\sum_{s,l,m,n} e^{i\Delta E t}\langle s|\alpha_{1}^{\dagger} | l \rangle\langle l|\alpha_{L}^\dagger| m\rangle \\ \nonumber
&\times &
\langle m|\alpha_1| n\rangle \langle n|\alpha_{L}| s\rangle,
\end{eqnarray}
where $\Delta E=E_{s}+E_{m}-E_{l}- E_{n}$ and $s,l,m,n$ are energy eigenstate indices.  In the long time limit, the contribution from all terms $\Delta E \neq 0$ vanishes due to the averaging over all eigenstates.  

If there exist left/right strong zero modes $\alpha_{l,r}$ in some parameter regimes,  then the whole spectra of the parafermion  Hamiltonian should be three-fold degenerated up to exponentially small finite-size corrections.  In other words,  the spectrum can be classified into triplets of eigenstates with different parity $P=1,\omega,\omega^2$,  that become exponentially degenerate as the system size increase.  For each eigenstate,  acting $\alpha_{l,r}$ on it cycles its parity by $\omega$.  Donating the three eigenstates in the subspace as $|\psi_{0,1,2}\rangle$,  with the index marking the parity.  In a suitable guage,  we assume $\alpha_l |\psi_2\rangle =  |\psi_1\rangle $,  $\alpha_l |\psi_1\rangle =  |\psi_0\rangle $ and $\alpha_l |\psi_0\rangle =  |\psi_2\rangle$. Due to the commutation relation $\alpha_l \alpha_r = \omega \alpha_r \alpha_l$,  the action of $\alpha_r$ should satisfy $\alpha_r |\psi_2\rangle =  \omega
 |\psi_1\rangle $,  $\alpha_r |\psi_1\rangle =  |\psi_0\rangle $ and $\alpha_r |\psi_0\rangle =  \omega^2|\psi_2\rangle$.  
 
\begin{figure}
  \centering\includegraphics[width=0.5\textwidth, height=7cm]{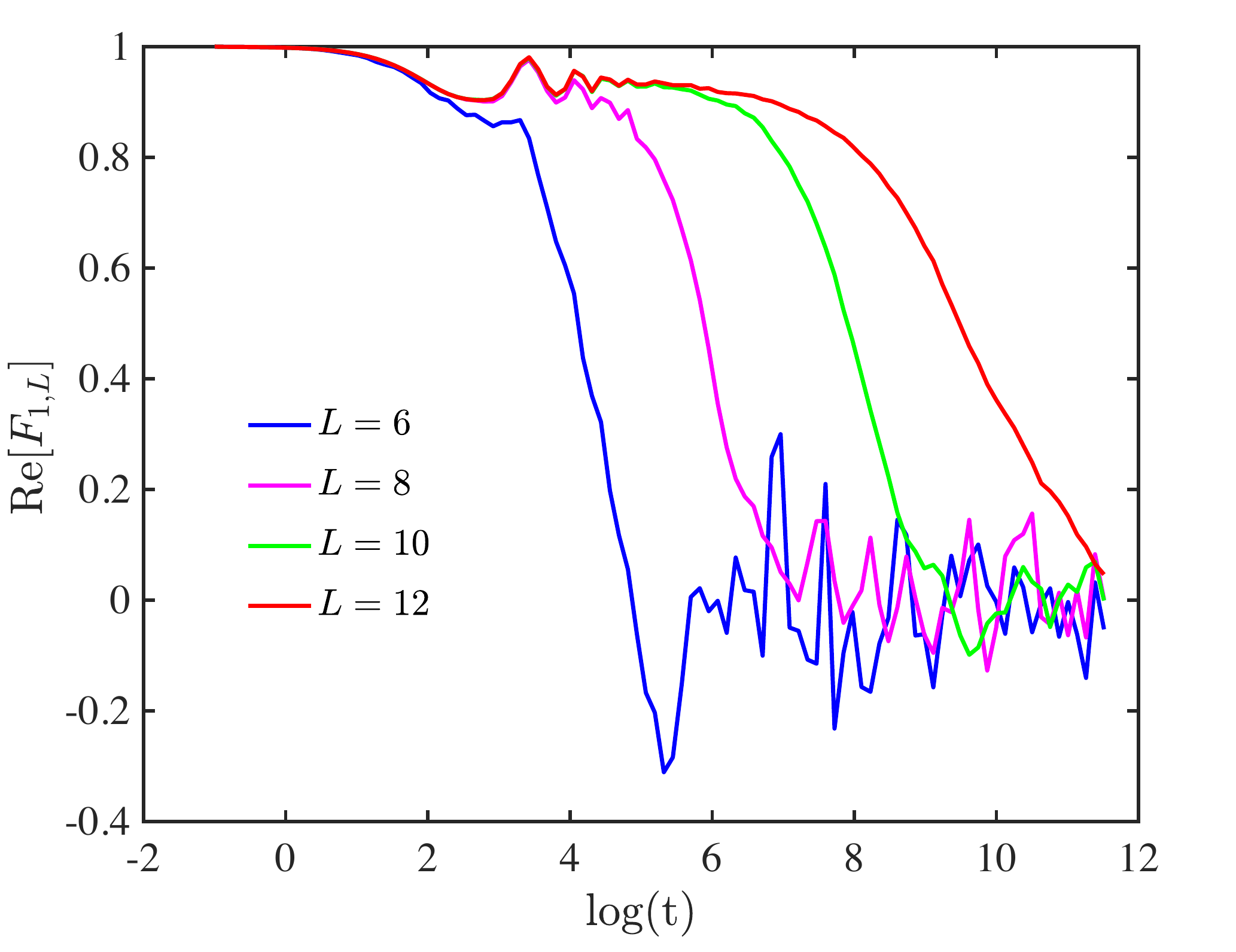}
  \caption{The OTOC ${\rm Re}[F_{1,L}(t)]$ between parafermions in  the two open ends  for  increasing chain length $L=6,8,10,12$ with ED method .  The parameter is $J_2=0.2, t_\text{tol} = 10^5$ and the time is in the log scale.}
  \label{figS5}
\end{figure}

In the topological limit $J_2 \rightarrow 0$,   $\alpha_l=\alpha_1$ and $\alpha_r=\alpha_{L}$,  all the non-diagonal components in the Eq.\ref{eq:decom} vanish and only the diagonal terms contribute,  in which case $E_s=E_l=E_m=E_n$.  When departing from this limit,  one expects $\alpha_{1,L}$ consists of a large overlap with $\alpha_{l,r}$ and the diagonal terms dominate the contribution.
To show the evidence of strong zero modes,  we calculate the OTOC ${\rm Re}[F_{1,L}(t)]$ between parafermions at the two open ends using the exact diagonalization method, see Fig. \ref{figS5}.  As the length of the chain increases, we find that  the scrambling time increases expotentionally in the regime $J_2=0.2$, which implies the existence of strong zero modes.

\end{document}